\documentstyle[prl,aps]{revtex}
\def\<{\langle}
\def\>{\rangle}
\def\d{\partial}
\def\x{{\bf x}}
\def\+{\dagger}
\def\YM{{\rm YM}}
\def\U1A{U(1)$_{\rm A}$}
\def\N=4{${\cal N}=4$}

\begin{document}
\twocolumn[\hsize\textwidth\columnwidth\hsize\csname@twocolumnfalse\endcsname
\preprint{CU-TP-xxx}
\title{Shear Viscosity of Strongly Coupled \N=4 Supersymmetric 
Yang-Mills Plasma}
\author{G.~Policastro$^{1,2}$, D.T.~Son$^{3,4}$, and A.O.~Starinets$^1$}
\address{$^1$Department of Physics, New York University, New York, New York
10003}
\address{$^2$Scuola Normale Superiore, Piazza dei Cavalieri 7, 56100,
Pisa, Italy}
\address{$^3$Physics Department, Columbia University, New York, New York 
10027}
\address{$^4$RIKEN-BNL Research Center, Brookhaven National Laboratory,
Upton, New York 11973}
\date{April 2001}
\maketitle
\begin{abstract}
Using the anti--de Sitter/conformal field theory correspondence, we
relate the shear viscosity $\eta$ of the finite-temperature \N=4
supersymmetric Yang-Mills theory in the large $N$, strong-coupling
regime with the absorption cross section of low-energy gravitons by a
near-extremal black three-brane.  We show that in the limit of zero
frequency this cross section coincides with the area of the horizon.
From this result we find $\eta={\pi\over8}N^2T^3$.  We conjecture that
for finite 't Hooft coupling $g_\YM^2N$ the shear viscosity is
$\eta=f(g_\YM^2N)N^2T^3$, where $f(x)$ is a monotonic function that
decreases from ${\cal O}(x^{-2}\ln^{-1}(1/x))$ at small $x$ to $\pi/8$
when $x\to\infty$.
\end{abstract}
\vskip 2pc]

{\em Introduction}.---At finite temperatures, the large distance, long
time behavior of gauge theories is described, as in any other fluid,
by a hydrodynamic theory \cite{LL}.  To write down the hydrodynamic
equations one has to know the thermodynamics (i.e., the equation of
state) of the medium, as well as the transport coefficients: the shear
and the bulk viscosities, the electrical conductivity [in the presence
of a U(1) gauge group], and the diffusion constants (in the presence
of conserved global charges).  Knowledge of these quantities in hot gauge
theories is crucial for numerous applications, the most notable
of which belong to the physics of the electroweak phase transition
in the early Universe \cite{Rubakov} and of the quark-gluon plasma
possibly created in heavy-ion collisions \cite{heavyion}.

When the gauge coupling is small (which requires, in the case of QCD,
temperatures much larger than the confinement scale), both the
equation of state and the transport coefficients are calculable
perturbatively.  At strong coupling (i.e., at temperatures not much
larger than the confinement scale), 
thermodynamics can be found nonperturbatively by lattice simulations,
but transport coefficients are beyond the reach of all current
numerical techniques.  This situation is very unfortunate, since the
quark-gluon plasma one hopes to create in 
heavy-ion experiments has relatively low temperature at which the
perturbation theory works very poorly.

Lacking methods to reliably compute the transport coefficients of
finite-temperature QCD, one should try to gain insight into models where
these coefficients can be computed nonperturbatively.  Recently,
powerful techniques based on the anti--de Sitter/conformal field
theory (AdS/CFT) correspondence have been developed, establishing, in
particular, the connection between the \N=4 supersymmetric Yang-Mills
(SYM) theory in the large coupling, large $N$ limit and classical
ten-dimensional gravity on the background of black three-branes
\cite{Maldacena,GKP,Witten1,report}.  This allows one to perform
analytical calculations in a strongly coupled four-dimensional gauge
theory.

In this Letter, we compute the shear viscosity $\eta$ of the strongly
coupled finite-temperature \N=4 SYM theory (the bulk viscosity of this
theory vanishes due to scale invariance).  We first relate, using
previously known results from the AdS/CFT correspondence, the shear
viscosity with the absorption cross section of low-energy gravitons
falling perpendicularly onto near-extremal black three-branes.  We
further show that this cross section is equal to the the area of the
horizon, in a way very similar to the case of black holes \cite{DGM}.
These facts provide enough information for us to find that
$\eta={\pi\over8}N^2T^3$, provided both the 't Hooft coupling and $N$
are large.  Remarkably, the shear viscosity approaches a constant
value in the large 't Hooft coupling limit, and this value is related
to the area of the horizon of the black brane.

{\em The viscosity}.---To start our discussion, we briefly review the
notion of viscosity in the context of finite-temperature field theory.
Consider a plasma slightly out of equilibrium, so that there is local
thermal equilibrium everywhere but the temperature and the mean
velocity slowly vary in space.  We define, at any point, the local
rest frame as the one where the three-momentum density vanishes:
$T_{0i}=0$.  The stress tensor, in this frame, is given by the
constitutive relation,
\begin{eqnarray}
  T_{ij} &=& \delta_{ij}p - \eta \biggl(\d_i u_j + \d_j u_i - 
  {2\over3} \delta_{ij} \d_k u_k\biggr)\nonumber\\ 
  & & - \zeta \delta_{ij} \d_k u_k
  \, ,
\end{eqnarray}
where $u_i$ is the flow velocity, $p$ is the pressure, and $\eta$ and
$\zeta$ are, by definition, the shear and bulk viscosities
respectively.  In conformal field theories like the \N=4 SYM theory,
the energy momentum tensor is traceless, ${T^\mu}_\mu=0$, so
$\varepsilon\equiv T_{00}=3p$ and the bulk viscosity vanishes
identically, $\zeta=0$.

All kinetic coefficients can be expressed, through Kubo relations, as
the correlation functions of the corresponding currents \cite{Hosoya}.
For the shear viscosity, the correlator is that of the stress tensor,
\begin{eqnarray}
  \eta &=& \lim_{\omega\to0}{1\over2\omega}\int\!dt\,d\x\, e^{i\omega t}
   \< [T_{xy}(t, \x), T_{xy}(0, 0)] \> \nonumber\\
  &=& \lim_{\omega\to0}
  {1\over2\omega i}[G_{\rm A}(\omega)-G_{\rm R}(\omega)]\, ,
  \label{Kubo}
\end{eqnarray}
where the average $\<\ldots\>$ is taken in the equilibrium thermal
ensemble, and $G_{\rm A}$ and $G_{\rm R}$ are the advanced and retarded
Green functions of $T_{xy}$, respectively.  In Eq.\ (\ref{Kubo}), 
the Green functions are computed at zero spatial momentum.
Though Eq.\ (\ref{Kubo}) can, in principle, be used to
compute the viscosity in weakly coupled field theories, this direct
method is usually very cumbersome, since it requires resummation of an
infinite series of Feynman graphs.  This calculation has been
explicitly carried out only for scalar theories \cite{Jeon}.  A more
practical method is to use the kinetic Boltzmann equation, which gives
the same results as the diagrammatic approach \cite{JeonYaffe}.


For gauge theories at weak coupling, $g^2N\ll1$, where throughout this
paper $g\equiv g_\YM$ is the gauge coupling, the shear viscosity has
the following parametric behavior,
\begin{equation}
  \eta = C {N^2T^3\over (g^2N)^2\ln(1/g^2N)} \, .
  \label{weak}
\end{equation}
Basically, $\eta$ is proportional to the product of the energy density
$\varepsilon\sim N^2T^4$ and the transport mean free time $\tau\sim
[(g^2N)^2\ln(1/g^2N)T]^{-1}$.  The numerical coefficient $C$ in Eq.\
(\ref{weak}), in principle, can be computed by solving the linearized
Boltzmann equation \cite{AMY}.

{\em The relation to graviton absorption}.---The key observation
underlying this work is that the right hand side of the Kubo formula
(\ref{Kubo}) is known to be proportional to the classical absorption
cross section of gravitons by black three-branes \cite{Klebanov,GKT}.
For completeness, we recall here the basic argument leading to this
correspondence.  Consider, in type IIB string theory, a configuration
of $N$ D3-branes stacked on top of each other.  The low-energy theory
living on the branes is the \N=4 U($N$) SYM theory.  On the other
hand, if $N$ is large, the stack of D3-branes has large tension, which
curves space-time.  In the limit of large 't Hooft coupling $g^2N$,
the three-brane geometry has small curvature and can be described by
supergravity.  Therefore, we have two descriptions of the same physics
in terms of strongly coupled gauge theory on the branes and classical
gravity on a certain background.

If one sends a graviton to the brane, there is some probability that
it will be absorbed.  On the gravity side, the absorption cross
section can be calculated by solving the wave equation on the
background metric.  On the gauge theory side, the rate of graviton
absorption measures the imaginary part of the stress tensor--stress
tensor correlator, since gravitons polarized parallel to the brane are
coupled to the stress-energy tensor of the degrees of freedom on the
brane.  The relation between the absorption cross section
$\sigma(\omega)$ of a graviton with energy $\omega$, polarized
parallel to the brane (say, along the $xy$ directions) and falling at a
right angle on the brane is related to the correlator in field theory
as \cite{Klebanov,GKT}
\begin{equation}
  \sigma(\omega) = {\kappa^2\over\omega} \int\!dt\,d\x\, e^{i\omega t}
   \< [T_{xy}(t, \x), T_{xy}(0, 0)] \> \, ,
  \label{sigmaT}
\end{equation}
where $\kappa=\sqrt{8\pi G}$, $G$ being the ten-dimensional
gravitational constant.  The relation (\ref{sigmaT}) has been
explicitly verified for zero-temperature field theory (or extremal
black branes, in the gravity language) \cite{Klebanov,GKT}.  Such a
check is possible because there is a nonrenormalization theorem for
the correlator of the stress-energy tensor \cite{GubserKlebanov}.

At finite temperature $T$, Eq.\ (\ref{sigmaT}) relates the graviton
absorption cross section by a near-extremal black brane having the
Hawking temperature equal to $T$ with a correlator in the hot SYM
theory \cite{Witten}.  Since there is no supersymmetry and no
nonrenormalization theorem is known to work at finite temperature,
one cannot explicitly verify the relation (\ref{sigmaT}).  We instead
view Eq.\ (\ref{sigmaT}) as a prediction of theory.  In particular,
taking the $\omega\to0$ limit one can relate $\sigma(\omega=0)$ with
the shear viscosity of hot SYM plasma,
\begin{equation}
  \eta = {1\over 2\kappa^2} \sigma(0) \, .
  \label{etasigma}
\end{equation}
Equation (\ref{etasigma}) implies that, for nonextremal black branes,
the graviton absorption cross section must not vanish in the limit of
zero frequency (in contrast to the extremal case where
$\sigma(\omega)\sim\omega^3$ at small $\omega$ \cite{Klebanov,GKT}),
and, by computing the zero-frequency value of $\sigma$ one obtains the
shear viscosity of the hot SYM plasma.  The problem of computing the
shear viscosity is now reduced to a problem of classical gravity.

The metric of a nonextremal black three-brane has the form
\cite{HorowitzStrominger,DuffLu}
\begin{eqnarray}
  ds^2 &=& H^{-1/2}(r) [-f(r)dt^2 + d\x^2] \nonumber\\
  & & +H^{1/2}(r) [f^{-1}(r) dr^2 + r^2 d\Omega_5^2] \, ,
  \label{metric}
\end{eqnarray}
where $H(r)=1+R^4/r^4$ and $f(r)=1-r_0^4/r^4$.  The extremal case
corresponds to $r_0=0$; the limit relevant for us is the near-extremal
one, $r_0\ll R$.  This metric has a horizon at $r=r_0$.  From the
existence of this horizon one should expect $\sigma(0)$ to be
nonvanishing.  Running ahead, we will show, by solving the wave
equation on the metric (\ref{metric}), that $\sigma(0)$ is {\em equal}
to the area of the horizon,
\begin{equation}
  \sigma(0) = \pi^3 r_0^3 R^2
  \label{sigma0}
\end{equation}
(the numerical coefficient $\pi^3$ is simply the area of the unit
five-sphere).  Using the formula for the Hawking temperature of the
metric (\ref{metric}),
\begin{equation}
  T = {r_0\over\pi R^2} \, ,
  \label{T}
\end{equation}
and the relation between $R$ and $N$ which is obtained by identifying
the Arnowitt-Deser-Misner mass per unit volume of the three-brane with
the tension of a stack of $N$ D3-branes \cite{Klebanov},
\begin{equation}
  R^4 = {\kappa N\over2\pi^{5/2}} \, ,
\end{equation}
we find the shear viscosity to be
\begin{equation}
  \eta = {\pi\over8} N^2 T^3 \, .
  \label{result}
\end{equation}
This is the main result of the paper.  Up to a constant, the shear
viscosity is equal to the entropy density \cite{entropy,entropy1}.
Both quantities are proportional to the area of the horizon.

{\em Solution to the radial wave equation}.---Now let us show how Eq.\
(\ref{sigma0}) is obtained.  We have to solve the $s$-wave radial
equation for a minimally coupled scalar (such as the graviton
polarized parallel to the brane),
$\d_\mu(\sqrt{-g}g^{\mu\nu}\d_\nu\phi)=0$.  In the metric
(\ref{metric}) this equation acquires the form
\begin{equation}
  \phi'' + {5r^4-r_0^4\over r(r^4-r_0^4)}\phi'
  +\omega^2 {r^4(r^4+R^4)\over(r^4-r_0^4)^2}\phi=0 \, .
  \label{radial1}
\end{equation}
The method we use to solve Eq.\ (\ref{radial1}) is the same matching
method that was used in the extremal case \cite{Klebanov,GKT}.  Since
ultimately we are interested in the limit $\omega\to0$, we will assume
$\omega\ll T$.  More details about our method, as well as the solution
to the radial equation in the opposite limit $\omega\gg T$ and for
higher partial waves, can be found in Ref.\ \cite{PS}.  Earlier
attempts to compute the absorption rate by nonextremal black branes
were made in Refs.\ \cite{earlier}.  As in the extremal case, we
search for the solution in several regions and match the result
wherever the regions overlap.  Let us go from small $r$ to large $r$,
starting from the horizon $r=r_0$.  The first region is the one just
outside the horizon: $r>r_0$, $r-r_0\ll r_0$.  In this case Eq.\
(\ref{radial1}) has the form
\begin{equation}
  \phi'' + {\phi'\over r-r_0} + {\lambda^2\over16}{\phi\over(r-r_0)^2}
  = 0 \, ,
  \label{phihor}
\end{equation}
where $\lambda=\omega/(\pi T)\ll1$.  The solution to Eq.\ (\ref{phihor}) is
\begin{equation}
  \phi = A \biggl( 1 - {r_0\over r}\biggr)^{-i\lambda/4} \, ,
  \label{phi1}
\end{equation}
where we have chosen the sign of the exponent so that the solution
corresponds to an incoming wave at the horizon.  When $\lambda$ is
small, (\ref{phi1}) is basically a constant, $A$, except for an
exponentially small region near $r_0$.  In the next region,
$r_0<r\ll\omega^{-1}$ (excluding $r$ exponentially closed to $r_0$),
the term proportional to $\omega^2$ in the left-hand side of Eq.\
(\ref{radial1}) can be dropped.  Indeed,
\begin{equation}
  \omega^2 r^8(r^4-r_0^4)^{-2} \ll (r-r_0)^{-2}
\end{equation}
due to $r\ll\omega^{-1}$, and 
\begin{equation}
  \omega^2r^4R^4(r^4-r_0^4)^{-2}\ll (r-r_0)^{-2}
\end{equation}
since $\omega\ll T\sim r_0R^{-2}$.  Equation (\ref{radial1}) now has
the form
\begin{equation}
  \phi'' + {5r^4 - r_0^4\over r(r^4-r_0^4)}\phi' =0 \, ,
\end{equation}
which possesses a trivial solution,
\begin{equation}
  \phi = A \label{phi2} \, ,
\end{equation}
which matches smoothly with Eq.\ (\ref{phi1}).  Finally, in the
outermost region, $r\gg R\gg r_0$, Eq.\ (\ref{radial1}) is simplified
to
\begin{equation}
  {d^2\phi\over dr^2} + {5\over r}{d\phi\over dr} + \omega^2\phi = 0
  \, ,
\end{equation}
which can be solved in terms of the Bessel functions,
\begin{equation}
  \phi(r) = \alpha{J_2(\omega r)\over (\omega r)^2} +
            \beta{Y_2(\omega r)\over (\omega r)^2}, \qquad r\gg R \, .
  \label{phi3}
\end{equation}
The regimes of validity of Eq.\ (\ref{phi2}), $r_0<r\ll\omega^{-1}$,
and of Eq.\ (\ref{phi3}), $r\gg R$, has an overlap since
$\omega^{-1}\gg R$ (this is the consequence of $R\gg r_0$ and
$\omega\ll T$).  In order for Eq.\ (\ref{phi3}) to match with Eq.\
(\ref{phi2}) in the overlapping region, one should require
\begin{equation}
  \alpha = 8A, \qquad \beta = 0 \, .
\end{equation}
The field at large distances can be decomposed into an incoming wave
and an outgoing wave,
\begin{equation}
  \phi(r) = 4A \biggl[ {H_2^{(1)}(\omega r)\over(\omega r)^2} +
                       {H_2^{(2)}(\omega r)\over(\omega r)^2} \biggr]
                       \, .
  \label{inout}
\end{equation}

The absorption probability $P$ is the ratio of the flux at $r=r_0$
from Eq.\ (\ref{phi1}) and the flux from the incoming wave in Eq.\
(\ref{inout}).  We find
\begin{equation}
  P = {\pi\over32} \omega^5 r_0^3 R^2 \, .
\end{equation}
Since the absorption cross section $\sigma$ is related to $P$ by
\cite{DGM}
\begin{equation}
  \sigma = {32\pi^2\over\omega^5}P \, ,
\end{equation}
we arrive to Eq.\ (\ref{sigma0}), which coincides with the area of the
horizon.  This is very similar to the universal result for black holes
\cite{DGM}.

Notice that in deriving Eq.\ (\ref{sigma0}) we require $\omega$ to be
much smaller than the Hawking temperature.  The absorption cross
section will deviate substantially from the zero-frequency limit if
$\omega$ is of order $T$.  In particular, the next correction to Eq.\
(\ref{sigma0}) is of order $\omega^2/T^2$ with a computable
coefficient \cite{PS}.

{\em Discussion}.---We have shown that the shear viscosity can be
computed in the strongly coupled \N=4 SYM theory from the AdS/CFT
correspondence.  Now let us try to interpret the result
(\ref{result}).  The power of $T$ in $\eta$ is completely fixed by the
dimensionality of $\eta$ and the scale invariance of the theory.  The
factor $N^2$ apparently comes from the number of degrees of freedom in
the plasma.  It is remarkable that the shear viscosity approaches a
constant value as one sends the 't Hooft coupling to infinity.  From
the relation $\eta\sim\varepsilon\tau$, one can interpret this
behavior as the indication that the ``relaxation time'' $\tau$ remains
of order $T^{-1}$ (but not much smaller) in the strong coupling limit.
Since the inverse relaxation time is comparable to the energy per degree
of freedom, the strongly coupled plasma cannot be viewed as a
collection of particles, and the formula $\eta\sim\varepsilon\tau$
does not applies in the strict sense.  However, one should expect that
the counting of the powers of $N$ still works.  This counting,
combined with the expressions for $\eta$ is the weak-coupling [Eq.\
(\ref{weak})] and strong-coupling [Eq.\ (\ref{result})] limits,
suggests that for finite 't Hooft coupling $g^2N$ the shear viscosity
has the form
\begin{equation}
  \eta = f(g^2N) N^2 T^3 \, ,
\end{equation}
where $f(x)\sim x^{-2}\ln^{-1}(1/x)$ when $x\ll1$ and $f(x)=\pi/8$
when $x\gg1$.  It is most likely that $f(x)$ is a monotonic
function of $x$.  One way to verify this conjecture is to compute
the ${\cal O}(1/g^2N)$ correction to $\eta$ in the strong coupling
limit.  If $f(x)$ is monotonic, then this correction must be
positive.  This is analogous to the behavior of the free energy
\cite{entropy1}, except that at small coupling $\eta\to\infty$, while
the free energy remains finite.

Recalling that $\sigma(\omega)$ deviates substantially from
$\sigma(0)$ when $\omega\sim T$, we see that the hydrodynamic theory
can describe processes occuring during times much larger than
$T^{-1}$, but breaks down for those whose typical time scale is of
order or less than $T^{-1}$.  One also should expect hydrodynamics to
work at spatial distances much larger than $T^{-1}$, but not at
distances of order or less than $T^{-1}$.  This is consistent with
$T^{-1}$ playing the role of the relaxation time in the limit
$g^2N\to\infty$.  There is apparently no separation of scales in the
strong coupling regime that would make a kinetic description possible:
$T^{-1}$ is the only time/length scale.  Thus, the viscosity $\eta$
cannot be computed from a Boltzmann-type equation.

In this paper, we have confined our attention to the most important
transport coefficient --- the shear viscosity.  As mentioned above,
the bulk viscosity vanishes identically due to the exact scale
invariance of the ${\cal N}=4$ SYM theory.  It would be useful to
compute other transport coefficients in this theory (for example, the
diffusion constant of the $R$ charges) at strong coupling using the
AdS/CFT correspondence.

We thank M.~Porrati for discussions.  G.P. is supported, in part, by the
Fondazione A.\ Della Riccia.  D.T.S. thank RIKEN, Brookhaven National
Laboratory, and U.S.\ Department of Energy [DE-AC02-98CH10886] for
providing the facilities essential for the completion of this work.
The work of D.T.S is supported, in part, by
a DOE OJI grant.

\end{document}